# Current-Controlled Negative Differential Resistance due to Joule Heating in TiO$_2$

A. S. Alexandrov,[1,2] A. M. Bratkovsky,[2*] B. Bridle,[1] S. E. Savel'ev,[1] D. B. Strukov[3] and R. Stanley Williams[2]

[1]*Department of Physics, Loughborough University, Loughborough LE11 3TU, United Kingdom*
[2]*Hewlett-Packard Laboratories, 1501 Page Mill Road, Palo Alto, California 94304*
[3]*Department of Electrical and Computer Engineering, University of California Santa Barbara Santa Barbara, CA 93106*
*\* To whom all correspondence should be sent, alex.bratkovski@hp.com*

## Abstract

We show that Joule heating causes current-controlled negative differential resistance (CC-NDR) in TiO$_2$ by constructing an analytical model of the voltage-current V(I) characteristic based on polaronic transport for Ohm's Law and Newton's Law of Cooling, and fitting this model to experimental data. This threshold switching is the 'soft breakdown' observed during electroforming of TiO$_2$ and other transition-metal-oxide based memristors, as well as a precursor to 'ON' or 'SET' switching of unipolar memristors from their high to their low resistance states. The shape of the V(I) curve is a sensitive indicator of the nature of the polaronic conduction.

The effects of Joule heating are very important in the operation of a wide variety of electronic devices, especially as they shrink to nanoscale sizes and the power densities become large. This has become especially apparent for the class of memristors based on resistance switching in transition metal oxides [1-11]. For these systems, there is often an initial electroforming step that is used to turn an inert capacitor with a fixed resistance >1G$\Omega$ into a memristor with a continuum of resistance states <10M$\Omega$, see e.g., extensive reviews in Ref. [12-14] . This process has been studied by several investigators, and it is thought to be initiated by a purely electrical conduction process, or soft breakdown [15,16], followed by the electrochemical reduction of oxide anions to form oxygen gas, which escapes the system, and oxygen vacancies in the oxide matrix [17]. The changes in conductivity of the oxide material induced by the movement of these vacancies, under the influence of an electric field, a concentration gradient or a temperature gradient, are responsible for the memristive switching [12, 13, 18, 19, 9, 14, 11].

There have been previous examinations of this heating, and models for the thermally assisted conduction have been proposed [15, 1]. Here we present a compact analytical model for the electrical conduction based on feedback between adiabatic polaron transport and Joule heating that allows us to fit the initial stages of experimental electroforming voltage-current V(I) characteristics of a nanoscale memristor in order to determine the temperature of the heated channel in the device as a function of current. This analysis provides both a thermometer inside the device and insight into the nature of the polaronic transport in $TiO_2$.

As-fabricated devices have a metal-oxide-metal structure in which successive layers are deposited by a physical or chemical vapor deposition process. The films can be either polycrystalline or amorphous, and the surfaces and resulting interfaces have some level of roughness. This in turn means that the oxide film has variations in thickness and local structure, which provides preferential sites for current to flow when a potential is applied across the electrodes. Thus, there are variations in the resulting V(I) characteristics from device to device caused by the fact that thickness, composition and structural differences will act as nucleating sites for the flow of current. The width of the current channel is usually 100nm or less [20], and there is usually only one such operating channel, even if the physical device is larger than 100μm$^2$. When a voltage is first applied, the current is small because the resistance is high. However, a threshold is reached at which the current increases rapidly, even though the voltage decreases. This Current-Controlled Negative Differential Resistance (CC-NDR) is also known as S-NDR (because of the shape of the I vs. V curve) and threshold switching. In the case of electroforming, the NDR is not reversible since the process yields oxygen vacancies that will then lead to memristive switching [17]. However, similar CC-NDR V(I) behavior is observed as the precursor to reversible ON or SET switching in unipolar and bipolar memristors [21, 22, 23], and is responsible for heating that then drives or assists the resistive switching.

In all these cases, the heating is not only rapid, but leads to very large temperature *gradients* in the oxide that can dramatically influence vacancy and other defect distributions. The diffusion of vacancies in the crystal field depends exponentially on temperature, so that variations of the local temperature could have a drastic effect on the distributions. A self-consistent molecular-dynamics type of simulation of local heating at the nanoscale might run into the fundamental problem of assigning a distinct temperature to each lattice unit cell, which is hard to justify from a thermodynamic standpoint [24].

In the absence of a rigorous approach to this fundamental problem, we consider here the effect of *homogeneous* heating of a memristor-type device on its current-voltage characteristics, applying Ohm's law $I = V/R(T)$ and Newton's Law of Cooling as they have been successfully applied to describe the nonlinear I-V characteristics of superconducting mesas [25] and thermometry of conducting channels in electroformed memristors [2, 5]. The temperature of a thin active layer is thus given by $T = T_0 + \alpha IV$, where $T_0$ is the bath temperature and $\alpha$ is the heat transfer coefficient or thermal resistance. The electronic resistance in oxide films is well described by the thermal activation law for small polaron mobility,

$$R(T) = \beta T^n \exp(E_a/k_B T), \qquad (1)$$

with the activation energy $E_a$, a constant $\beta$ and $k_B$ the Boltzmann constant. The power $n$ depends on the adiabatic versus nonadibatic behavior of the polarons, $n = 1$ or $n = 3/2$, respectively, which is still ill determined for $TiO_2$ [26]. Using Ohm's law we can construct V(I) curves by solving the transcendental equation

$$i = v/(t + iv)^n \exp[-1/(t+iv)]. \qquad (2)$$

Here $i = I[\alpha\beta/(E_a/k_B)^{1-n}]^{1/2}$ and $v = V[\alpha/\beta(E_a/k_B)^{1+n}]^{1/2}$ are dimensionless current and voltage, respectively, and $t = k_B T_0/E_a$ is the reduced bath temperature. Eq. (2) can be written in a parametric form as:

$$v = [p(t+p)^n]^{1/2}\exp[1/2(t+p)], \qquad (3a)$$

$$i = [p(t+p)^{-n}]^{1/2}\exp[-1/2(t+p)], \qquad (3b)$$

where $p = iv$. We can thus compute both $v$ and $i$ as functions of the dimensionless power $p$ as a parameter, and combine these values to plot the V(I) characteristic as shown in Figure 1. Independent of the model parameters, $\alpha$ and $\beta$, V(I) has a local maximum at sufficiently low reduced bath temperatures t<0.25, which is the signature characteristic of threshold switching.

In the case of Fig. 1, we have fit Eq. (3) to experimental data from Ref. [27] collected during the electroforming of a nanoscale $TiO_2$-based crosspoint memristor at room temperature. In this case, the 1.5 nm Ti adhesion layer, 8 nm Pt bottom electrode and the 11 nm Pt top electrode were deposited by electron-beam evaporation at room temperature. The 50 nm thick Ti dioxide film was deposited by sputtering from a $TiO_2$ target in 3 mTorr Ar and 550K substrate temperature. The 50 nm bottom and top electrodes were patterned by ultraviolet-nanoimprint lithography to define the crosspoint junction. The oxide was a blanket layer without patterning. The Ti adhesion layer diffused through the thin Pt bottom electrode during the deposition of the $TiO_2$ layer and partially reduced the oxide film at the bottom interface to form a good electrical contact to the device [27].

We see that the fit shown in Fig. 1 is excellent for currents up to 0.04 mA, a reduced bath temperature t=0.11, and exponent $n = 1$. This corresponds to polaronic conduction in the adiabatic regime, in agreement with the transport observed for bulk $TiO_2$ [26]. For larger currents, the data deviates from the fit because of the existence of a second local maximum in V(I), which is the onset of irreversible behavior in the device. The corresponding local temperature at this point is about 600-650K, cf. Fig.1.

Using the absolute values of the voltage and current at the first local maximum in the V(I) curve in Fig.1, 2.53 V and I=6.67µA, respectively, and converting to the dimensionless values v=2.4, i=0.008 and using t=0.11 for the relative bath temperature, we estimate $E_a$=214 meV, the thermal coefficient $\alpha=3.2\times10^6$ K/W and the resistance coefficient $\beta$=0.49 Ohm/K. Our estimate of the polaron activation energy is very close to the value (248 meV) measured in bulk films of $TiO_2$ [26]. With our value of $\alpha$, we can estimate the internal temperature of the heated channel for any current, as shown in Fig. 1, with the first local maximum in V(I) occurring at about 350K.

The pre-exponential factor in the conductivity is $\sigma_0 = L/(A\beta)$, where $L$ is the thickness and $A$ the cross-section area of the active part of the device (conducting channel that is forming). In the adiabatic regime [see Ref. 26, Eq.(9)]:

$$\sigma_0 = \frac{Ne^2a^2\nu}{k_B}, \tag{4}$$

where $N$ is the polaron density, $a$ the hopping range and $\nu$ the optical phonon frequency. Estimating the average hopping range as the mean distance between $O^{-2}$ vacancies, $a \approx (N/2)^{-1/3}$, and assuming $L$=10nm, $A$=(50nm)$^2$, and $\nu$=1.1×10$^{13}$ Hz [26], we obtain $N = 4[k_B\sigma_0/(e^2\nu)]^3 \approx 3\times10^{20}$ cm$^{-3}$. The value of L used here is an order of magnitude estimate that is less than the physical thickness of the deposited film because it represents the narrowest region of the oxide after it has been reduced by the adhesion layer. Since the crosspoint area is small, our approximations should be acceptable. The resulting carrier density $N$ is reasonable for a sputtered oxide film and well in line with previous observations of perovskite oxide films with concentrations of vacancies near high work function electrodes reaching a few atomic percent [28].

The forming step in TiO$_2$ thus starts with a `soft' breakdown, revealed by the CC-NDR, that electronically heats the material, which then leads to an exponentially enhanced drift of oxygen anions that in turn leads to a thermal runaway. In TiO$_2$, the electrochemical reduction of the anions occurred at a bias voltage ≥2V, and thus an estimated field of ~0.2 GV/m, which created a large concentration of vacancies and a metallic suboxide conducting channel [17]. This `soft breakdown' preserved the reversible bipolar resistive switching of the devices, and should be contrasted with the usual dielectric breakdown in oxide films that leads to an irreversible shorting of a thin film [29,30]. Local temperatures reaching 650K lead to the crystallization of amorphous TiO$_2$ to form anatase [9,11]. This phase transition may be the origin of the second NDR peak in the experimental V(I) curve in Figure 1, in analogy with the CC-NDR resulting from the insulator-to-metal phase transition described by Pickett et al. [23].

We illustrate the importance of the relative bath temperature $t = k_B T_0/E_a$ in Fig. 2, which is on the order of 0.1 for $TiO_2$ at room temperature. We see that V(I) curves depend strongly on $t$ and $n$. This shows that collecting V(I) curves in the Joule-heating CC-NDR regime of crosspoint devices before irreversible changes occur at a variety of temperatures is a very sensitive means for studying polaronic transport in transition-metal-oxide thin films, and that fitting to our model provides a direct route for determining the activation energy for polaron hopping and the $n$ parameter, which is otherwise difficult to extract experimentally.

During the past several decades, many materials with a high density of carriers and low mobility on the order or even less than the Ioffe-Regel limit (~1cm$^2$/V.s) have been discovered [31]. There has been some confusion in the literature, particularly about the use of the terms "large" and "small" polarons and bipolarons. Some authors define the `large polaron' as a mobile electron moving together with a self-induced extended polarization well in ionic crystals, and the `small polaron' as an immobile object completely localized within a unit cell (site). However, from the pioneering work of Holstein [32] and others it is known that small polarons can propagate via coherent tunneling in a narrow band due to translational symmetry of the lattice at temperatures below the Debye temperature, while at higher temperatures the transport can proceed via thermally activated hopping. Some distinct features of small polarons include a low mobility, coherent band motion at sufficiently low temperatures, activated mobility at higher temperatures, and a mid-infrared maximum in optical conductivity. Our results agree with the main conclusions of the comprehensive studies of small polarons in rutile by Bogomolov *et al.* [33], whose experimental investigations of the drift and Hall mobilities, the infrared absorption, the thermoelectric power, and comparison with small polaron theory provided convincing evidence for small polaron transport in this material [32, 31].

Understanding CC-NDR due to Joule heating in $TiO_2$ thin films is not only important technologically for preparing and operating nano-memristors, but it also provides an excellent opportunity for the further understanding of small polaron properties. Our analytical model based on polaron conduction and Newton's Law of Cooling explains the CC-NDR that represents the initial stage of electroforming in $TiO_2$ memristors as well as the heating required for the SET operation of a unipolar memristor, and also provides an internal thermometer that can be used to determine the internal temperature of the device up to the limit of applicability of the model.

**Acknowledgments**

We thank J. Joshua Yang for providing the experimental data shown in Fig. 1, and helpful discussions, and John Paul Strachan for the detailed discussions of his experiments.

Figure 1

A fit of experimental V(I) electroforming data from Ref. [27] (blue dots) to Eq.(3) for a 50 nm × 50 nm crosspoint $TiO_2$-based memristor. The red solid line is the fit with $n = 1$, which corresponds to polaronic transport in the adiabatic regime, and the dashed line is the fit with $n = 3/2$, which is for the nonadiabatic regime, see Ref.[26]. The superior fit for $n = 1$ suggests the adiabatic regime. Also shown by the dashed-dotted line is the ramp in local heating, $(T-T_0)/T_0$, where $T_0$ is the bath temperature. The abrupt deviation of the experimental V(I) data from the fitting curve at I=0.065 mA represents the onset of irreversible changes in the device, possibly the crystallization of amorphous $TiO_2$ to form anatase, which occurs at a local temperature of about 600-650K .

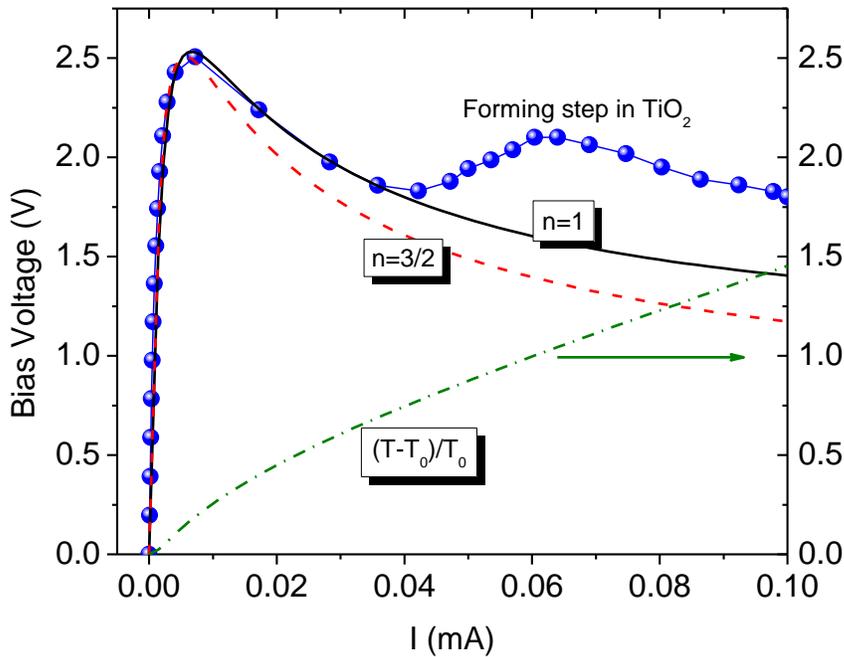

Figure 2.

Effect of the relative bath temperature $t = k_B T_0/E_a$ on the V(I) characteristics of polaronic conductors. For TiO$_2$, $t \approx 0.1$, and thus this system displays CC-NDR in the initial stages of electroforming.

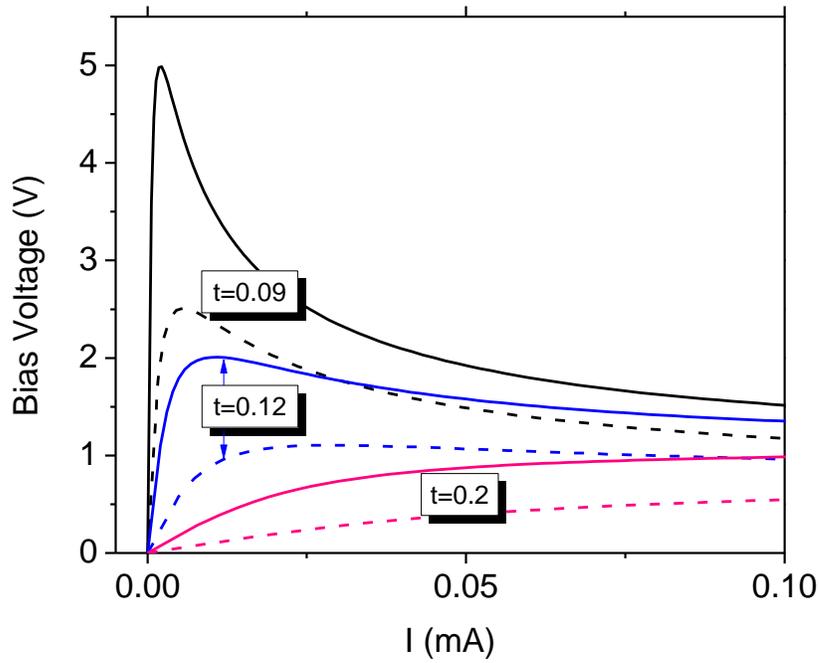